\documentclass[dvips]{article}
\usepackage{graphicx}
\usepackage{amssymb}
\usepackage{graphicx}
\usepackage{dcolumn}
\usepackage{amsmath}
\usepackage{lscape}
\usepackage{longtable}

\begin{document}
\textwidth 16cm
\newcommand{\bd}{\begin{document}}
\newcommand{\ed}{\end{document}}
\newcommand{\bc}{\begin{center}}
\newcommand{\ec}{\end{center}}
\newcommand{\bfr}{\begin{flushright}}
\newcommand{\efr}{\end{flushright}}
\newcommand{\lt}{\left}
\newcommand{\rt}{\right}
\newcommand{\vs}{\vspace}
\newcommand{\hs}{\hspace}
\newcommand{\beq}{\begin{equation}}
\newcommand{\eeq}{\end{equation}}
\newcommand{\lb}{\linebreak}
\newcommand{\pb}{\pagebreak}
\newcommand{\mb}{\makebox}
\newcommand{\fb}{\framebox}
\newcommand{\mc}{\multicolumn}
\newcommand{\ben}{\begin{enumerate}}
\newcommand{\een}{\end{enumerate}}
\newcommand{\bit}{\begin{itemize}}
\newcommand{\eit}{\end{itemize}}
\newcommand{\ol}{\overline}
\newcommand{\un}{\underline}
\newcommand{\lefq}{\lefteqn}
\newcommand{\ba}{\begin{array}}
\newcommand{\ea}{\end{array}}
\newcommand{\beqa}{\begin{eqnarray}}
\newcommand{\eeqa}{\end{eqnarray}}
\newcommand{\beqas}{\begin{eqnarray*}}
\newcommand{\eeqas}{\end{eqnarray*}}
\newcommand{\bfg}{\begin{figure}}
\newcommand{\efg}{\end{figure}}
\newcommand{\bds}{\begin{displaymath}}
\newcommand{\eds}{\end{displaymath}}
\newcommand{\btb}{\begin{tabbing}}
\newcommand{\etb}{\end{tabbing}}
\newcommand{\para}{\parallel}
\newcommand{\pad}{\partial}
\newcommand{\nn}{\nonumber}
\newcommand{\la}{\leftarrow}
\newcommand{\ra}{\rightarrow}
\newcommand{\lgla}{\longleftarrow}
\newcommand{\lgra}{\longrightarrow}
\newcommand{\La}{\Leftarrow}\newcommand{\Ra}{\Rightarrow}
\newcommand{\Lra}{\Leftrightarrow}
\newcommand{\Lgla}{\Longleftarrow}
\newcommand{\Lgra}{\Longrightarrow}
\newcommand{\bm}{\boldmath}
\newcommand{\lan}{\langle}
\newcommand{\ran}{\rangle}
\renewcommand{\a}{\alpha}
\renewcommand{\b}{\beta}
\newcommand{\g}{\gamma}
\newcommand{\G}{\Gamma}
\renewcommand{\d}{\delta}
\newcommand{\eps}{\epsilon}
\newcommand{\Th}{\Theta}
\newcommand{\s}{\sigma}
\newcommand{\lam}{\lambda}
\newcommand{\D}{\Delta}
\newcommand{\vare}{\varepsilon}
\newcommand{\pr}{\prime}
\newcommand{\ro}{\rho}
\newcommand{\nab}{\nabla}
\newcommand{\m}{\mu}
\newcommand{\n}{\nu}
\newcommand{\Sg}{\Sigma}
\newcommand{\p}{\pi}
\newcommand{\R}{I\!\!R}
\newcommand{\om}{\omega}
\newcommand{\Om}{\Omega}
\newcommand{\ze}{\zeta}
\newcommand{\vart}{\vartheta}
\newcommand{\tri}{\triangle}
\newcommand{\f}{\frac}
\newcommand{\iny}{\infty}
\newcommand{\pro}{\propto}
\bc {\huge Quantum isotonic nonlinear oscillator as a Hermitian counterpart of Swanson Hamiltonian and pseudo-supersymmetry } \ec

\vs{1cm}

\bc
{\it \"Ozlem Ye\c{s}ilta\c{s}{\footnote {e-mail : yesiltas@gazi.edu.tr}\\
Department of Physics, Faculty of Science,
Gazi University,
06500 Ankara, Turkey\\
\vspace{.15cm}

}} \ec
\vs{2cm}
\begin{abstract}
Within the ideas of pseudo-supersymmetry, we have studied a non-Hermitian Hamiltonian $H_{-}=\omega(\xi^{\dag} \xi+\frac{1}{2})+\alpha \xi^{2}+\beta \xi^{\dag 2}$, where $\alpha \neq \beta$ and $\xi$ is a first order differential operator, to obtain the partner potentials $V_{+}(x)$ and $V_{-}(x)$ which are new isotonic and isotonic nonlinear oscillators, respectively, as  the Hermitian equivalents of  the non-Hermitian partner Hamiltonians $H_{\pm}$. We have provided an algebraic way to obtain the spectrum and wavefunctions of a nonlinear isotonic oscillator. The solutions of $V_{-}(x)$ which are Hermitian counterparts of Swanson Hamiltonian are obtained under some parameter restrictions that are found. Also, we have checked that if the intertwining operator satisfies $\eta_{1} H_{-}=H_{+} \eta_{1}$, where $\eta_{1}=\rho^{-1} \mathcal{A} \rho$ and $\mathcal{A}$ is the first order differential operator, which factorizes Hermitian equivalents of $H_{\pm}$.
\end{abstract}

\newpage
\section{Introduction}
A class of quantum potentials known as conditionally exactly solvable(CES)  models which include one or more specific fixed coupling constants was introduced by Dutra \cite{dutra}. Since then, CES models have attracted considerable attention \cite{gro, Junk,calo, dong, Carinena, fellow, ses, Kra}. Recently, a non-polynomial, one dimensional isotonic nonlinear oscillator potential \cite{Carinena, fellow, ses, Kra} and its generalization to a $d$ dimensional model have been studied \cite{HSY}. On the other hand, specific symmetries of the isotonic
potential have been investigated \cite{xx}. Previously, a CES model which is an isotonic nonlinear oscillator potential constructed by a superpotential in \cite{Junk}. The isotonic potential which is a harmonic oscillator with a barrier potential is given by \cite{calo,dong}
\begin{equation}\label{iss}
    V(x)=\frac{1}{2}x^{2}+\frac{1}{2}g x^{-2}
\end{equation}
where $g> -\frac{1}{2}$. Classically, isotonic potential model has a property which is the family of periodic solutions with the same period \cite{clas} and quantum mechanically the energy spectrum is equidistant \cite{weiss}.
Recently, for the Schr\"{o}dinger equation, Cari\~{n}ena et al investigated the solutions of
\begin{equation}\label{schr?}
    U(x)=x^{2}+8\frac{2x^{2}-1}{(2x^{2}+1)^{2}}
\end{equation}
where the new additional term is non-polynomial and it is the sum of two centripetal barriers in the complex plane  \cite{Carinena}. Later Fellows and Smith showed that this potential is a supersymmetric partner of a harmonic oscillator \cite{fellow}. Within position dependent mass aproach, it is shown that the source and the target potentials share the same energy where the isotonic potential is the reference potential \cite{Kra}. At the same time, the fundamental framework of pseudo-Hermitian quantum mechanics has been proposed in \cite{mos} and the idea of non-Hermitian Hamiltonians has spread to many branches of physics such as electrodynamics \cite{elo}, biological physics \cite{bio}, quantum information \cite{info} and theoretical physics \cite{the,bqr,Roy,Y}.

In this study, we discuss the isotonic oscillator with a non-polynomial term which is a Hermitian counterpart of non-Hermitian Swanson Hamiltonian \cite{swanson,q1,af1,af2,cher,musumbu,Bagchitanaka,Proy}, in other words, our model can be a generalization of the non-Hermitian models for a nonlinear isotonic potential within the framework of pseudo-supersymmetry \cite{ir,das,roych}. We shall also seek for the solutions of the corresponding system. The plan of the paper is as follows. In section 2, we introduce  a brief review for factorization and hierarchy of Hamiltonians. Section 3 devoted to non-Hermitian partner Hamiltonians $H_{\pm}$ which are isospectral  except for the ground state. Pseudo- supersymmetry relations between the partner Hamiltonians are discussed. In section 4, the solutions of the Hermitian equivalent Hamiltonians of $H_{\pm}$ are introduced.

The aim of this work devotes us to use factorization approach. As is well known, the superalgebra constructed by the super Hamiltonian $\mathcal{H}_{s}$ and the supercharges $Q$, $Q^{\dag}$ is \cite{cooper}
\begin{equation}\label{sa}
    \mathcal{H}_{s}=\{Q,Q^{\dag}\}
\end{equation}
where $Q$, $Q^{\dag}$ commute with $\mathcal{H}_{s}$ and '$\dag$' is the Hermitian conjugation:
\begin{equation}\label{Hs}
    Q^{2}=Q^{\dag2}=[Q, \mathcal{H}_{s}]=[Q^{\dag}, \mathcal{H}_{s}]=0.
\end{equation}
The supercharges are given by the following matrices,
\begin{equation}\label{qq}
    Q=\begin{pmatrix}
        0 & 0 \\
        A & 0 \\
      \end{pmatrix},~~
      ~~~Q^{\dag}=\begin{pmatrix}
                      0 & A^{\dag} \\
                      0 & 0 \\
                    \end{pmatrix}.
\end{equation}
Here, $\mathcal{H}_{s}$ is diagonal that can be factorized in the manner
\begin{equation}
\mathcal{H}_{s}=\begin{pmatrix}
  \mathcal{H}_{+} & 0 \\
  0 & \mathcal{H}_{-} \\
\end{pmatrix},~~~~
\mathcal{H}_{-}=A^{\dag}A,~~ \mathcal{H}_{+}=AA^{\dag}
\end{equation}
where $A$ and $A^{\dag}$ are first order differential operators given by a more general realization \cite{plas}
\begin{equation}\label{aA}
    A=a(x)\frac{d}{dx}+b(x), ~~~~A^{\dag}=-a(x)\frac{d}{dx}+b(x)-a^{\prime}(x), ~~~~^{\prime}=\frac{d}{dx}
\end{equation}
where $b(x)$ is the superpotential related to the partner potentials
\begin{equation}\label{part}
    U_{+}(x)=U_{-}(x)+2b^{\prime}(x)a(x)-a(x)a^{\prime\prime}(x), ~~~
\end{equation}
which are elements of the partner Hamiltonians:
\begin{equation}\label{ph}
    \mathcal{H}_{\pm}=-\frac{d}{dx}a^{2}(x)\frac{d}{dx}+U_{\pm}(x).
\end{equation}
The ground states can be introduced as
\begin{equation}\label{gs}
    A\psi^{-}_{0}=0 ~~or ~~A^{\dag}\psi^{+}_{0}=0
\end{equation}
where $\psi^{\pm}_{n}$ ($n=0, 1,...$) stands for the bound state wavefunctions for $\mathcal{H}_{\pm}$. Let $E^{\pm}_{n}$ are the energy eigenvalues of $\mathcal{H}_{\pm}$, then we suppose that the $\mathcal{H}_{-}$ is known and has normalizable vanishing ground state energy in the unbroken SUSY case. In this manner, we have
\begin{equation}\label{kn}
    \psi^{-}_{n}(x)=(E^{+}_{n})^{-\frac{1}{2}} A^{\dag}\psi^{+}_{n}, ~~~~\psi^{+}_{n}=(E^{-}_{n})^{-\frac{1}{2}} A\psi^{-}_{n},~~~E^{+}_{n}=E^{-}_{n+1}, ~~~E^{-}_{0}=0.
\end{equation}
\section{Non-Hermitian partner Hamiltonians}
In order to describe a physical system, a non-Hermitian Hamiltonian $H_{-}$ should be $\eta$-pseudo Hermitian \cite{mos},
 \begin{equation}\label{etah}
    H^{\dag}_{-}=\eta H_{-} \eta^{-1}
 \end{equation}
 where $\eta$ is a linear, invertible operator. If $\eta$ is not a Hermitian operator, then the Hamiltonian is known as weakly pseudo-Hermitian \cite{solo}. To find a Hermitian Hamiltonian $h_{-}$, one can use a similarity transformation $h_{-}=\rho H_{-} \rho^{-1}$ and the metric operator is obtained using $\eta=\rho^{2}$. Consider the ladder operators $\xi$, $\xi^{\dag}$ that are first order differential operators
\begin{equation}\label{xi}
    \xi=a(x)\frac{d}{dx}+b(x),~~~~~~\xi^{\dag}=-a(x)\frac{d}{dx}+b(x)-a^{\prime}(x)
\end{equation}
which will be used in a pseudo-Hermitian quadratic Hamiltonian which is known as Swanson model \cite{swanson}:
\begin{equation}\label{sw}
    H_{-}=\omega(\xi^{\dag} \xi+\frac{1}{2})+\alpha \xi^{2}+\beta \xi^{\dag 2},~~~~\alpha \neq \beta.
\end{equation}
where parameters $\omega, \alpha, \beta \in \mathbb{R}$ and it was shown that the eigenvalues are real and positive $\omega^{2}-4 \alpha \beta >0$ when $\xi$ represents linear harmonic oscillator annihilation operator. Now, (\ref{sw}) can be cast in a differential operator form:
\begin{equation}\label{sH}
    H_{-}=-\bar{\omega}\frac{d}{dx}a^{2}(x)\frac{d}{dx}+b_{1}(x)\frac{d}{dx}+c_{1}(x)
\end{equation}
where $\bar{\omega}=\omega-\alpha-\beta$, $\bar{\omega}>0$. One gives $b_{1}(x)$ and $c_{2}(x)$ as \cite{Bagchitanaka}
\begin{equation}\label{b1}
   b_{1}(x)= (\alpha-\beta)a(x)(2b(x)-a^{'}(x))
\end{equation}
and
\begin{equation}\label{c2}
\begin{split}
    c_{1}(x) &=(\omega+\alpha+\beta)b^{2}(x)-(\omega+2\beta) a^{'}(x)b(x)-(\omega-\alpha+\beta)a(x)b^{'}(x)\\+& \beta(a(x)a^{''}(x)+a^{'2}(x))
   -\delta a^{'}(x)+\frac{\omega}{2}.
  \end{split}
\end{equation}
Hermitian equivalent of (\ref{sw}) which is $h_{-}$
\begin{equation}\label{her}
    h_{-}=\rho H_{-} \rho^{-1}
\end{equation}
obtained by using $\rho=\exp\left(-\frac{1}{2\bar{\omega}}\int dx \frac{b_{1}(x)}{a^{2}(x)}\right)$, then $h_{-}$ takes the form \cite{bqr, Bagchitanaka}
\begin{equation}\label{eff}
    h_{-}=-\bar{\omega}\frac{d}{dx} a^{2}(x)\frac{d}{dx}+V_{-}(x)
\end{equation}
where \cite{Bagchitanaka}
\begin{equation}\label{V1}
\begin{split}
    V_{-}(x)&=\left(\frac{(\alpha-\beta)^{2}}{\bar{\omega}}+\bar{\omega}+2(\alpha+\beta)\right)b(x)(b(x)-a^{'}(x))-(\bar{\omega}+\alpha+\beta)a(x)b^{'}(x)\\
    &+ \frac{\alpha+\beta}{2}a(x)a^{''}(x)+\frac{1}{4}\left(\frac{(\alpha-\beta)^{2}}{\bar{\omega}}+2(\alpha+\beta)\right)a^{'2}(x)+
     \frac{\bar{\omega}+\alpha+\beta}{2}.
    \end{split}
\end{equation}
In order to factorize $h_{-}$, it is written as a product of two operators
\begin{equation}\label{f1}
    h_{-}=\mathcal{A}^{\dag}\mathcal{A}.
\end{equation}
Let us give an ansatze for the operators we use in above
\begin{equation}\label{a,b}
   \mathcal{A}=\sqrt{\bar{\omega}}a(x)\frac{d}{dx}+\tilde{b}(x),~~\mathcal{A}^{\dag}=-\sqrt{\bar{\omega}}a(x)\frac{d}{dx}+\tilde{b}(x)-
   \sqrt{\bar{\omega}}a^{\prime}(x).
\end{equation}
where $\tilde{b}(x)$ and $a(x)$ are real and they can be chosen to generate $V_{\pm}(x)$. Thus, $V_{-}(x)$ becomes
\begin{equation}\label{Vek}
    V_{-}(x)=\tilde{b}^{2}(x)-\sqrt{\bar{\omega}}\frac{d}{dx}(a(x)\tilde{b}(x)).
\end{equation}
As is well known, $h_{-}$ is isospectral to its partner Hamiltonian $h_{+}$ that is
\begin{equation}\label{h1}
    h_{+}=\mathcal{A} \mathcal{A}^{\dag}
\end{equation}
and it can be  expressed as
\begin{equation}\label{h2}
    h_{+}=-\bar{\omega}\frac{d}{dx} a^{2}(x)\frac{d}{dx}+V_{+}(x).
\end{equation}
Now, the partner potential of $V_{-}(x)$ is
\begin{equation}\label{V2}
    V_{+}(x)=\tilde{b}^{2}(x)+\sqrt{\bar{\omega}}\left(-\sqrt{\bar{\omega}}a(x)a(x)^{\prime\prime}+a(x)\tilde{b}^{\prime}(x)-\tilde{b}(x) a^{\prime}(x)\right).
\end{equation}
Exploiting the fact that the non Hermitian partner Hamiltonians $H_{\pm} $ and their Hermitian counterparts $h_{\pm}$ are related by a similarity transformation
\begin{equation}\label{cp}
    H_{\pm}=\rho^{-1} h_{\pm} \rho,
\end{equation}
then we can get $H_{+}$ as
\begin{equation}\label{H2}
    H_{+}=-\bar{\omega}\frac{d}{dx}a^{2}(x)\frac{d}{dx}+b_{1}(x)\frac{d}{dx}+\frac{b^{\prime}_{1}}{2}-\frac{b_{1}^{2}}{4\bar{\omega}a^{2}}+
   V_{+}(x).
\end{equation}
Finally, we can give a diagramatic representation for a chain of Hamiltonians we used  as follows
\begin{equation}\label{diag}
    H_{-}\rightarrow h_{-}\rightarrow h_{+}\rightarrow H_{+}
\end{equation}
such that we can arrive at partner Hamiltonian $H_{+}$  using this chain.
\subsection{Factorization of the Hermitian counterparts of Swanson Hamiltonian}
Some choices of $a(x)$ and $b(x)$ may lead to an effective potential model. In \cite{q1, Bagchitanaka}, it was shown that when the commutator is constant, i.e $[\xi, \xi^{\dag}]=1$, then the relation between $a(x)$ and $b(x)$ can be given as \cite{Bagchitanaka}
\begin{equation}\label{commut}
    b(x)=\frac{a^{'}(x)}{2}+\int \frac{dx}{2a(x)}.
\end{equation}
The more general case corresponds to $[\xi, \xi^{\dag}]\neq 1$. Then, let us use the ansatz for $a(x)$ and $b(x)$
\begin{eqnarray}
  a(x) &=& x^{2} \\
  b(x) &=& \frac{1}{x}+\frac{c x}{x^{2}+d}
\end{eqnarray}
in (\ref{V1}), then $V_{-}(x)$ becomes
\begin{eqnarray}\label{V11}
    V_{-}(x)=\frac{a_{1}}{x^{2}}+2(a_{3}+2a_{4})x^{2}+(-2a_{1}+a_{2})(c+1)+a_{5}\\+
    c\frac{(2a_{1}+a_{1}c+2a_{1}d-3a_{2}d)x^{2}+2a_{1}(1+d)-a_{2}d}{(x^{2}+d)^{2}} \nonumber
\end{eqnarray}
where $c, d$ are some real constants. Here $a_{i}, ~~i=1,2...5$ can be recalled from (\ref{V1}), i.e:
\begin{eqnarray}\label{CC}
  a_{1} &=& \frac{{(\alpha-\beta)^{2}}}{\bar{\omega}}+\bar{\omega}+2(\alpha+\beta),~~~~~~ a_{2} = \bar{\omega}+\alpha+\beta,~~~~~~a_{3} = \frac{\alpha+\beta}{2}\\
  a_{4} ~&= & \frac{1}{4}\left(\frac{{(\alpha-\beta)^{2}}}{\bar{\omega}}+2(\alpha+\beta)\right),~~~~~~ a_{5} = \frac{\bar{\omega}+\alpha+\beta}{2}.
\end{eqnarray}
With an eye on what is to follow, let us first match (\ref{V11}) with (\ref{Vek}) to factorize $h_{-}$, so another ansatz is introduced for $\tilde{b}(x)$ which is,
\begin{equation}\label{bt}
    \tilde{b}(x)=\frac{\mu}{x}-\varrho x+\frac{\lambda x}{x^{2}+d}.
\end{equation}
At that case, $V_{-}(x)$ is written as
 \begin{equation}\label{Va}
    V_{-}(x)= \frac{\mu^{2}}{x^{2}}+\varrho(\varrho+3\sqrt{\bar{\omega}})x^{2}-\mu(2\varrho+\sqrt{\bar{\omega}})
    + \lambda \frac{-(2\varrho+\sqrt{\bar{\omega}})x^{4}+(-3d\sqrt{\bar{\omega}}+\lambda+2\mu-2d\varrho)x^{2}+2d\mu}{(x^{2}+d)^{2}}.
\end{equation}
Hence, we can give the coupling constants of $V_{-}(x)$ in (\ref{Va}) in terms of $\omega, \alpha, \beta$ so that we can compare (\ref{Va}) and (\ref{V11}). At the same time, its partner $V_{+}(x)$ becomes
\begin{equation}\label{Vb}
    V_{+}(x)=\frac{\mu^{2}}{x^{2}}+(\varrho^{2}+\varrho\sqrt{\bar{\omega}}-2\bar{\omega})x^{2}-\mu(2\varrho+3\sqrt{\bar{\omega}})+
    \lambda\frac{-(2\varrho+3\sqrt{\bar{\omega}})x^{4}+(\lambda+2\mu-2d\varrho-d\sqrt{\bar{\omega}}+2d\mu)x^{2}+2d\mu}{(d+x^{2})^{2}}.
\end{equation}
After some straightforward algebra, partner potentials happen to be
\begin{eqnarray}
  V_{-}(x) = \frac{\mu^{2}}{x^{2}}+\varrho(\varrho+3\sqrt{\bar{\omega}})x^{2}-(\mu+\lambda)(\sqrt{\bar{\omega}}+2\varrho)\\ \nonumber
  +\lambda\frac{(2\varrho d+2\mu+\lambda-d\sqrt{\bar{\omega}})x^{2}+d(2\mu+d\sqrt{\bar{\omega}}+2\varrho d)}{(x^{2}+d)^{2}} \\
  V_{+}(x) = \frac{\mu^{2}}{x^{2}}+\varrho(\varrho+\sqrt{\bar{\omega}}-2\bar{\omega})x^{2}-(\mu+\lambda)(3\sqrt{\bar{\omega}}+2\varrho) \nonumber
  \\ +\lambda\frac{(2\varrho d+2\mu+\lambda+5d\sqrt{\bar{\omega}})x^{2}+d(2\mu+3d\sqrt{\bar{\omega}}+2\varrho d)}{(x^{2}+d)^{2}}.
\end{eqnarray}
We seek for a solvable $V_{+}(x)$ which implies some parameter restrictions such that the non-polynomial part in $(38)$ vanishes within these conditions below
\begin{equation}\label{para}
\lambda=-2d\sqrt{\bar{\omega}}, ~~~~\mu= -\frac{d}{2}(2\varrho+3\sqrt{\bar{\omega}}).
\end{equation}
Herefrom, if we plug (\ref{para}) into $V_{+}(x)$, it turns into
\begin{equation}\label{son}
  V_{+}=  \frac{\mu^{2}}{x^{2}}+(\varrho^{2}+\varrho\sqrt{\bar{\omega}}-2\bar{\omega})x^{2}+
  2d\left(\varrho+\frac{7\sqrt{\bar{\omega}}}{2}\right)\left(\varrho+\frac{3\sqrt{\bar{\omega}}}{2}\right).
\end{equation}
The same procedure is followed for $V_{-}(x)$, it follows that
\begin{equation}\label{Vekk}
    V_{-}=\frac{\mu^{2}}{x^{2}}+\varrho(\varrho+3\sqrt{\bar{\omega}})x^{2}+2d\left(\varrho+\frac{7\sqrt{\bar{\omega}}}{2}\right)\left
    (\varrho+\frac{\sqrt{\bar{\omega}}}{2}\right)+
    4\bar{\omega}d^{2}\frac{3x^{2}+d}{(x^{2}+d)^{2}}.
\end{equation}
Comparing (\ref{V11}) and (\ref{Vekk}), as we stated before, it may give us a chance to write $\mu, \lambda, \varrho$ in terms of $\omega, \alpha, \beta$. Then,
\begin{eqnarray}\label{con2}
  a_{1} &=& \mu^{2}=\frac{d^{2}}{4}(2\varrho+3\sqrt{\bar{\omega}})^{2} \\ \label{con3}
  2(a_{3}+2a_{4}) &=& \varrho(\varrho+3\sqrt{\bar{\omega}}) \\
  (a_{2}-2a_{1})(c+1)+a_{5} &=& 2d(\varrho+\frac{7\sqrt{\bar{\omega}}}{2})(\varrho+\frac{\sqrt{\bar{\omega}}}{2}) \label{con4}
\end{eqnarray}
and
\begin{eqnarray}
  c(-3a_{2}d+2a_{1}+a_{1}c+2a_{1}d) &=& 12\bar{\omega}d^{2} \\
  2a_{1}(1+d)-a_{2}d &=& 4\bar{\omega}d^{3}.
\end{eqnarray}
We can solve $c$ using last two relations above:
\begin{equation}\label{c}
    c_{1,2}=\frac{1}{2a_{1}}\left(2a_{2}d-4\bar{\omega}d^{3}\pm \sqrt{(4\bar{\omega}d^{3}-2a_{2}d)^{2}+48a_{1}\bar{\omega}d^{2}}\right).
\end{equation}
Using (\ref{con2}), (\ref{con3}) and (\ref{con4}), $\varrho$ can be found as
\begin{equation}\label{rrh}
    \varrho=\frac{1}{2}(\pm \sqrt{4X-27\bar{\omega}}-4\sqrt{\bar{\omega}})
\end{equation}
where
\begin{equation}\label{xx}
    X=\frac{4a_{1}}{d^{2}}-8(a_{3}+2a_{4})+\frac{1}{2d}\left((a_{2}-2a_{1})(c+1)+a_{5}\right).
\end{equation}
There are some restrictions on parameters such as $\varrho>0$, so we take $\varrho=\frac{1}{2}( \sqrt{4X-27\bar{\omega}}-4\sqrt{\bar{\omega}})$. Inside the square root in (\ref{rrh}) must be positive, $\varrho$ is positive, then $4X > 43 \bar{\omega}$. Because $a_{2}-2a_{1}=-\frac{\bar{\omega}^{2}+3(\alpha+\beta)\bar{\omega}+2(\alpha-\beta)^{2}}{\bar{\omega}} <0$, then we need  $c<0, |c|>1$ and we use the negative sign for $c$ which is $c=\frac{1}{2a_{1}}(2a_{2}d-4\bar{\omega}d^{3}- \sqrt{(4\bar{\omega}d^{3}-2a_{2}d)^{2}+48a_{1}\bar{\omega}d^{2}})$. When (\ref{rrh}), (\ref{c}) are satisfied, then Hermitian counterpart of $H_{-}$ which is $h_{-}$ given by (\ref{eff}) is factorized.
\subsection{Pseudosupersymmetry}
Here, $H_{-}$ is diagonalizable with a discrete spectrum which admits a set of complete biorthonormal eigenvectors $\{|\Psi_{n}\rangle, |\Phi_{n}\rangle\}$:
\begin{eqnarray}
  H_{-} |\Psi_{n}\rangle &=& E_{n} |\Psi_{n}\rangle,~~~H_{-}^{\dag}|\Phi_{n}\rangle=E^{*}_{n} |\Phi_{n}\rangle\\
  \sum_{n} |\Phi_{n}\rangle \langle \Psi_{n}|&=& \sum_{n} |\Psi_{n}\rangle \langle \Phi_{n}|=1.
\end{eqnarray}
On the other hand, the intertwining relations between $h_{\pm}$ can be written as
\begin{equation}\label{eq1}
    h_{-}\mathcal{A}^{\dag}=\mathcal{A}^{\dag}h_{+}
\end{equation}
and
\begin{equation}\label{eq2}
    h_{+}\mathcal{A}=\mathcal{A}h_{-}.
\end{equation}
And, $H_{-}$ given in (\ref{sw}) is a non-Hermitian and diagonalizable Hamiltonian with real or complex conjugate eigenvalues. If there exists an operator $\eta_{1}$ such that
\begin{equation}\label{ps1}
    \eta_{1}H_{-}=H_{+}\eta_{1}
\end{equation}
where $H_{+}$ is the partner Hamiltonian of $H_{-}$, then, the intertwining operator $\eta_{1}$ is given by $\eta_{1}=\rho^{-1} \mathcal{A} \rho$ where $\mathcal{A}$ and its adjoint satisfy (\ref{eq1}) and (\ref{eq2}). This relation $\eta_{1}=\rho^{-1} \mathcal{A} \rho$ was proven before \cite{das}. The intertwining operator that links a non-Hermitian Hamiltonian to the adjoint of its pseudo-supersymmetric partner Hamiltonian was also studied \cite{roych}. If one uses the similarity transformation
\begin{equation}\label{rel}
    H_{+}=\rho^{-1} h_{+} \rho
\end{equation}
then, if we multiply (\ref{rel}) by $\rho^{-1}\mathcal{A}$ from right, we have:
\begin{equation}\label{ps2}
    H_{+}\rho^{-1} \mathcal{A}=\rho^{-1} h_{+} \mathcal{A}
\end{equation}
and if we use (\ref{eq2}) in (\ref{ps2}), we get
\begin{eqnarray}
    \rho^{-1} \mathcal{A} h_{-}&=& H_{+} \rho^{-1} \mathcal{A}\\
    \rho^{-1}\mathcal{A} \rho\rho^{-1}h_{-}\rho&=&H_{+}\rho^{-1}\mathcal{A}\rho \Rightarrow \\
    \rho^{-1} \mathcal{A}\rho H_{-}&=&H_{+} \rho^{-1} \mathcal{A}\rho.
\end{eqnarray}
Then, it is seen that the intertwining operator $\eta_{1}$ is given by $\eta_{1}=\rho^{-1} \mathcal{A} \rho$.
Following the same way we can obtain $\eta_{2}H_{+}=H_{-}\eta_{2}$ where $\eta_{2}=\rho^{-1}\mathcal{A}^{\dag}\rho$. Here, $\eta_{1}$ and its pseudo-adjoint $\eta_{2}$ can be shown as
\begin{equation}\label{adj}
    \eta^{\sharp}_{2}=\eta^{-1} \eta^{\dag}_{2} \eta=\rho^{-1} \mathcal{A} \rho=\eta_{1}
\end{equation}
which leads to construct pseudo- super algebra of non-Hermitian supersymmetry. Thus, the operators $Q$, $Q^{\sharp}$ become
\begin{equation}Q=
\begin{pmatrix}
  0 & \eta_{2} \\
  0 & 0 \\
\end{pmatrix},
~~Q^{\sharp}=\begin{pmatrix}
               0 & 0 \\
               \eta_{1} & 0 \\
             \end{pmatrix}.
\end{equation}
The pseudo- super Hamiltonian can be introduced as
\begin{equation}\label{hem}
    \mathcal{H}=\begin{pmatrix}
  H_{-} & 0 \\
  0 & H_{+} \\
\end{pmatrix}
\end{equation}
where pseudo- super charges satisfy
\begin{equation}\label{hhh}
    \mathcal{H}=[ Q,Q^{\sharp} ]_{+}=0.
\end{equation}
Now, let us find $\eta_1$  for our problem, it can be written as
\begin{equation}\label{et1}
    \eta_{1}=\sqrt{\bar{\omega}}x^{2}\frac{d}{dx}+\frac{1}{\sqrt{\bar{\omega}}x(d+x^{2})}((\alpha-\beta-\varrho\sqrt{\bar{\omega}})x^{4}+
    ((\alpha-\beta)(d-c-1)-(\frac{7d\bar{\omega}}{2}+2\varrho d\sqrt{\bar{\omega}}))x^{2}+d(\mu\sqrt{\bar{\omega}}-\alpha+\beta))
    \end{equation}
which connects two partner Hamiltonians $H_{\pm}$. We can arrive at $H_{+}$ using $\eta_{1}$ given above.
\section{Solutions}
In this section, we aim to obtain the solutions of the partner Hamiltonians with (\ref{son}) and (\ref{Vekk}). Let  eigenfunctions of $h_{\pm}$ and $\tilde{h}_{\pm}$ be $\psi^{\pm}(x)$ and $\tilde{\varphi}^{\pm}(x)$. Then, $h_{\pm}$ can be transformed into $\tilde{h}_{\pm}$,
\begin{equation}\label{hx}
    \tilde{h}_{\pm}=\theta h_{\pm}\theta^{-1}=-\bar{\omega}a^{2}(x)\frac{d^{2}}{dx^{2}}-\bar{\omega}a(x)a^{\prime}(x)\frac{d}{dx}+V_{\pm}(x)+\frac{\bar{\omega}}{4}a^{\prime 2}(x)+\frac{\bar{\omega}}{2}a(x)a^{\prime\prime}(x)
\end{equation}
where $\theta=\sqrt{a(x)}$. One may use the change of independent variable in (\ref{hx}) as,
\begin{equation}\label{iv}
    z=\int^{x} \frac{dy}{\sqrt{\bar{\omega}}a(y)}
\end{equation}
and for our problem it equals to $z=-\frac{1}{\sqrt{\bar{\omega}}x}$. Then we arrive at
\begin{equation}\label{har}
    \tilde{h}_{\pm}=-\frac{d^{2}}{dz^{2}}+\left(V^{\pm}(x)+\frac{\bar{\omega}}{4}a^{\prime 2}(x)+\frac{\bar{\omega}}{2}a(x)a^{\prime\prime}(x)\right)_{x\rightarrow z}
\end{equation}
where we can write partner potentials $\tilde{V}_{\pm}=V_{\pm}+\frac{\bar{\omega}}{4}a^{\prime 2}(x)+\frac{\bar{\omega}}{2}a(x)a^{\prime\prime}(x)$ in terms of $z$ as
\begin{equation}\label{vva}
    \tilde{V}_{+}(z)=\mu^{2}\bar{\omega} z^{2}+\frac{\varrho^{2}+\sqrt{\bar{\omega}}\varrho}{\bar{\omega}}\frac{1}{z^{2}}+2d(\varrho+\frac{7\sqrt{\bar{\omega}}}{2})
    (\varrho+\frac{3\sqrt{\bar{\omega}}}{2})
\end{equation}
and
\begin{equation}\label{vav}
    \tilde{V}_{-}(z)=\mu^{2}\bar{\omega} z^{2}+\frac{\varrho^{2}+3\sqrt{\bar{\omega}}\varrho+2\bar{\omega}}{\bar{\omega}}\frac{1}{z^{2}}+
    2d(\varrho+\frac{7\sqrt{\bar{\omega}}}{2})(\varrho+\frac{\sqrt{\bar{\omega}}}{2})+4\bar{\omega}d+
    4\bar{\omega} d\frac{2d\bar{\omega}z^{2}-1}{(d\bar{\omega}z^{2}+1)^{2}}.
\end{equation}
We know that (\ref{vva}) and (\ref{vav}) are isospectral. In \cite{RNK}, an eigenvalue equation which is given by
\begin{equation}\label{pott}
    \left(-\frac{d^{2}}{dz^{2}}+\frac{\emph{A}}{z^{2}}+\emph{B} z^{2}\right)\chi_{n}(z)=\epsilon_{n}\chi_{n}(z)
\end{equation}
has exact eigenvalues and eigenfunctions as
\begin{equation}\label{e1}
    \epsilon_{n}=2\delta(2n+\gamma),~~n=0,1,2,...
\end{equation}
where $\gamma=1+\frac{1}{2}\sqrt{1+4\emph{A}}$, $\delta=\sqrt{\emph{B}}$ and
\begin{equation}\label{e2}
    \chi_{n}(z)=N z^{\gamma-1/2} e^{-\delta z^{2}/2}~~ _{1}F_{1}(-n; \gamma; \delta z^{2}),~~~n=0,1,2,...
\end{equation}
This model was studied by Goldman and Krivchenkov\cite{gold} who showed that the energy spectrum of this potential is an infinite set of equidistant
energy levels. Then, if we use  $\hat{\omega}=\frac{d\sqrt{\bar{\omega}}}{2}(2\varrho+3\sqrt{\bar{\omega}})$ and $\gamma=\frac{\varrho+\sqrt{\bar{\omega}}}{\sqrt{\bar{\omega}}}+\frac{1}{2}$ for the potential $\tilde{V}_{+}(z)$, exact eigenvalues and eigenfunctions of $\tilde{V}_{+}(z)$ are given by
\begin{equation}\label{ene}
    E^{+}_{n}=2\hat{\omega} (2n+\frac{2\rho}{\sqrt{\bar{\omega}}}+5)
\end{equation}
and
\begin{equation}\label{phia}
    \tilde{\varphi}^{+}_{n}(z)=C_{n} z^{\gamma-\frac{1}{2}}  e^{-\frac{1}{2}\hat{\omega} z^{2}}~
     _{1}F_{1}(-n;\gamma;\hat{\omega}z^{2})
\end{equation}
where $C_{n}$ is the normalization constant which is given by \cite{RNK}
\begin{equation}\label{nor}
    C_{n}=(-1)^{n}\sqrt{\frac{2\hat{\omega}^{\gamma}(\gamma)_{n}}{n!\Gamma(\gamma)}}
\end{equation}
and the solutions $\tilde{\varphi}^{+}_{n}(z)$ are square integrable on $(0,\infty)$ \cite{RNK}. The relation between the confluent hypergeometric function  and Laguerre polynomials may be used later:
\begin{equation}\label{F}
    _{1}F_{1}(-n;b+1;y)=\frac{n!}{(b+1)_{n}}L^{b}_{n}(y).
\end{equation}
Here the Pochhammer symbol $(s)_{n}=\frac{\Gamma(s+n)}{\Gamma(s)}$ is used for both (\ref{nor}) and (\ref{F}). Now, in order to find the eigenfunctions of $\tilde{h}_{-}$ we can follow an algebraic way such that we can start with a mapping between $h_{\pm}$ and $\tilde{h}_{\pm}$ introduced before:
\begin{eqnarray}\label{h1}
    \tilde{h}_{\pm}&=&\theta h_{\pm} \theta^{-1}. \\ \nonumber
    \end{eqnarray}
Also, this relation can be written as
\begin{eqnarray}\label{h1}
  \tilde{h}_{+} &=& \theta \mathcal{A}\mathcal{A}^{\dag} \theta^{-1} \\
   &=& \left(\theta \mathcal{A} \theta^{-1}\right) \left( \theta \mathcal{A}^{\dag}\theta^{-1} \right) \nonumber \\
   &=&\tilde{\mathcal{A}} \tilde{\mathcal{A}}^{\dag} \nonumber
\end{eqnarray}
and following the same way one obtains
\begin{equation}\label{h2}
    \tilde{h}_{-}=\tilde{\mathcal{A}}^{\dag}\tilde{\mathcal{A}}
\end{equation}
where $\tilde{\mathcal{A}}$ and $\tilde{\mathcal{A}}^{\dag}$ can be introduced as
\begin{eqnarray}
  \mathcal{\tilde{A}} &=& \sqrt{\bar{\omega}} a(x)\frac{d}{dx}+\tilde{b}(x)-\frac{\sqrt{\bar{\omega}}}{2} a^{\prime}(x) \\
  \mathcal{\tilde{A}}^{\dag} &=& -\sqrt{\bar{\omega}} a(x)\frac{d}{dx}+\tilde{b}(x)-\frac{\sqrt{\bar{\omega}}}{2} a^{\prime}(x).
\end{eqnarray}
Finally we can use $\mathcal{\tilde{A}}^{\dag}$ to find $ \varphi^{-}_{n}(z)$ which is the solution of $\tilde{V}_{-}(z)$:
\begin{eqnarray}\label{psik}
    \tilde{\varphi}^{-}_{n}(z)&=&\tilde{\mathcal{A}}^{\dag} \tilde{\varphi}^{+}_{n}(z)=\left(-\sqrt{\bar{\omega}} a(x)\frac{d}{dx}+\tilde{b}(x)-\frac{\sqrt{\bar{\omega}}}{2} a^{\prime}(x)\right)_{x\rightarrow z} C^{'}_{n} z^{\gamma-1/2} e^{-\frac{1}{2}\hat{\omega}z^{2}} L^{\gamma-1}_{n}(\hat{\omega}z^{2})\\
    &=& \left(-\frac{d}{dz}+\hat{\omega}z+\frac{\varrho}{\sqrt{\bar{\omega}}}\frac{1}{z}+\frac{2d\bar{\omega} z}{1+d\bar{\omega}z^{2}}\right)C^{'}_{n} z^{\gamma-1/2} e^{-\frac{1}{2}\hat{\omega}z^{2}} L^{\gamma-1}_{n}(\hat{\omega}z^{2}) \label{psik1}
\end{eqnarray}
where we use $\mu$ and $\lambda$ from (\ref{para}) in (\ref{psik}). Using a relation \cite{HSY}
\begin{equation}\label{lag}
    \frac{d}{dt} L^{\beta}_{n}(t)=-L^{\beta+1}_{n-1}(t)
\end{equation}
and the identity of Laguerre polynomials
\begin{equation}\label{ide}
    L^{\beta}_{n}(t)=L^{\beta}_{n-1}+L^{\beta-1}_{n}(t^{})
\end{equation}
helps us to re-write $\varphi^{-}_{n}(z)$ as \cite{HSY}
\begin{equation}\label{fi}
    \tilde{\varphi}^{-}_{n}(z)=\frac{2 C^{'}_{n} z^{\gamma+1/2}e^{-\hat{\omega} z^{2}/2}}{z^{2}+(d\bar{\omega})^{-1}} [(\gamma+n+1)L^{\gamma-1}_{n}(\hat{\omega}z^{2})-
    (n+1)L^{\gamma-1}_{n+1}(\hat{\omega}z^{2})+\gamma L^{\gamma}_{n}(\hat{\omega}z^{2})].
\end{equation}
On the other hand, one can show that the solutions $\psi^{\pm}_{n}(z)$ can be normalized. For example, $\psi^{+}_{n}$ can be given as
\begin{equation}\label{ps+}
    \psi^{+}_{n}(z)=\mathcal{N} z^{\gamma+1/2} e^{-\frac{1}{2}\hat{\omega} z^{2}}~
     _{1}F_{1}(-n;\gamma;\hat{\omega}z^{2})
\end{equation}
where $\mathcal{N}$ is the normalization constant. To normalize $\psi^{+}_{n}$, we may introduce the integral \cite{abro}
\begin{equation}\label{int1}
    \int^{\infty}_{0}  u^{2v-1} e^{-j u^{2}}~ _{1}F_{1}(-n;v;ju^{2}) ~ _{1}F_{1}(-m;v;ju^{2}) du=\frac{1}{2} \frac{n! \Gamma(v)}{j^{v}(v)_{n}} \delta_{mn}
\end{equation}
for $v>0$, $m,n=0,1,2,...$ and $\delta_{mn}=0$ if $m\neq n$, $\delta_{mn}=1$ if $m=n$. Now we use
\begin{equation}\label{e1}
    _{1}F_{1}(-n;b;y)=\sum^{n}_{k=0} \frac{(-n)_{k} y^{k}}{(b)_{k}k!}
\end{equation}
then we have
\begin{equation}\label{e2}
   J_{mn}= \sum^{n}_{k=0} \sum_{l=0}^{m} \frac{(-m)_{l} (-n)_{k} \hat{\omega}^{k+l}}{(\gamma)_{k}(\gamma)_{l}k! l!}\int^{\infty}_{0} z^{2\gamma+2k+2l+1}e^{-\hat{\omega}z^{2}} dz.
\end{equation}
The integral representation of the gamma function is
\begin{equation}\label{gama}
    \Gamma(x)=\int^{\infty}_{0}e^{-t} t^{x-1} dt,~~~~x>0,
\end{equation}
and using a variable change, we obtain
\begin{equation}\label{e3}
    J_{mn}=\frac{1}{2\hat{\omega}^{\gamma+1}} \sum^{n}_{k=0} \left(\sum^{m}_{l=0} \frac{(-m)_{l} (\gamma+k+1)_{l}}{(\gamma)_{l} l!}\right) \frac{(-n)_{k} \Gamma(\gamma+k+1)}{(\gamma)_{k}k!}.
\end{equation}
The term in the bracket in above relation corresponds to hypergeometric function $_{2}F_{1}$, then we re-write $J_{mn}$ as
\begin{equation}\label{e4}
    J_{mn}=\frac{1}{2\hat{\omega}^{\gamma+1}} \sum^{n}_{k=0}~ _{2}F_{1}(-m;\gamma+k+1;\gamma;1) \frac{(-n)_{k} \Gamma(\gamma+k+1)}{(\gamma)_{k}k!}.
\end{equation}
Using Chu-Vandermonde identity which is given by \cite{RNK}
\begin{equation}\label{chw}
    _{2}F_{1}(a;b;c;1)=\frac{\Gamma(c) \Gamma(c-a-b)}{\Gamma(c-a)\Gamma(c-b)}
\end{equation}
in (\ref{e4}), we get
\begin{equation}\label{e5}
    J_{mn}=\frac{\Gamma(\gamma)}{2\hat{\omega}^{\gamma+1}} \sum^{n}_{k=0} \frac{(-n)_{k}\Gamma(k+\gamma+1)}{(\gamma)_{k}k!} \frac{\Gamma(m-k-1)}{\Gamma(\gamma+m)\Gamma(-k-1)}.
\end{equation}
Using the identity \cite{RNK}
\begin{equation}\label{e6}
    (-k)_{n}=\left\{
               \begin{array}{ll}
                 \frac{(-1)^{n}k!}{(k-n)!}, & \hbox{$0\leq n\leq k$;} \\
                 0, & \hbox{$n>k$.}
               \end{array}
             \right.
\end{equation}
and (\ref{e5}), we obtain
\begin{equation}\label{e7}
    J_{mn}=\frac{(n+\gamma)(n+1)!\Gamma(\gamma)}{2\hat{\omega}^{\gamma+1}(\gamma)_{n}}
\end{equation}
for $k=n=m$. Finally, normalization constant becomes
\begin{equation}\label{e8}
    \mathcal{N}=(-1)^{n} \sqrt{\frac{2\hat{\omega}^{\gamma+1}(\gamma)_{n}}{\bar{\omega}(n+\gamma)(n+1)!\Gamma(\gamma)}}.
\end{equation}

\section{Conclusion}
To conclude, we have derived a new class of isotonic nonlinear oscillator using the concepts of pseudo-supersymmetry. We have studied a generalized non-Hermitian Hamiltonian  and its partner that can generate solvable isotonic and  nonlinear oscillator potentials by using
an appropriate mapping, i.e. a similarity transformation. Hermitian partner Hamiltonians which include effective solvable new isotonic and its nonlinear partner potential are factorized and it is seen that the nonlinear oscillator is solvable under parameter restrictions given by (\ref{para}). We have also checked that the intertwining operator $\eta_{1}$ that connects non-Hermitian partner Hamiltonians $H_{\pm}$ can be given in terms of a transformation of the factorization operator $\mathcal{A}$ as $\eta_{1}=\rho^{-1}\mathcal{A}\rho$. At the end, we give solutions of isotonic nonlinear oscillator after a coordinate and eigenfunction transformations. We have seen that the eigenvalues are real and positive which agrees with the results of \cite{swanson}.  We introduce formulae for all the discrete eigenvalues and normalized eigenfunctions.

\medskip



\end{document}